\documentstyle[12pt,aaspp]{article}

\def\etal{{\it et~al.\ }}
\def\eg{{\it e.g.\ }}
\def\ie{{\it i.e.\ }}
\def\gtwid{\mathrel{\raise.3ex\hbox{$>$\kern-.75em\lower1ex\hbox{$\sim$}}}}
\def\ltwid{\mathrel{\raise.3ex\hbox{$<$\kern-.75em\lower1ex\hbox{$\sim$}}}}

\lefthead {Illarionov and Krolik} 
\righthead {} 

\begin{document}

\title{Self-Shielding of X-rays and $\gamma$-rays in Compact Sources}    

\author{Andrei F. Illarionov}

\affil{Space Telescope Science Institute, Baltimore MD and P.N. Lebedev Physical
Institute, Moscow}

\and

\author{Julian H. Krolik}

\affil{Department of Physics and Astronomy, Johns Hopkins University, 
	   Baltimore, MD 21218}

\begin{abstract}
     It is generally supposed that when the ``compactness" $l \equiv
L\sigma_T/(r m_e c^3)$ in photons above the pair-production threshold is
large, few $\gamma$-rays can escape.  We demonstrate that even when $l \gg 1$,
if the high energy and low energy photons are produced in
geometrically-separated regions, many of the $\gamma$-rays can, in fact,
escape. Pair-production along a thin surface separating the
two sources creates enough Compton optical depth to deflect most
of the low energy photons away from the high energy ones.
Those few low-energy photons which penetrate the shielding surface are reduced
in opacity by advection to large distance
and small density, by relativistic beaming along the inner edge of
the surface, and by Compton upscattering to higher energies inside the surface.
The pairs in this surface flow outward relativistically,
forming a structure resembling a pair-dominated mildly relativistic jet.
\end{abstract}

\section{Introduction}

    One of the most basic concepts in the study of astrophysical $\gamma$-ray
sources is that of ``compactness" (see, {\it e.g.}, the review by Svensson
1986).  Its importance stems from many causes,
but one of its central implications has to do with the optical depth of
a $\gamma$-ray source to pair production.  Two photons with energies
$x_1$ and $x_2$ (in units of $m_e c^2$) may react to produce
an $e^{\pm}$ pair when their energies satisfy the relation $x_1 x_2
\geq 2/(1 - \cos\theta)$, where $\theta$ is the angle between their
directions of motion.  If a luminosity $L$ in photons of energy
just above threshold is made isotropically within a source of size $R$,
then the optical depth to pair production $\tau_{\gamma\gamma}$ is
greater than or of order unity when
\begin{equation}
l \equiv {L \sigma_T \over R m_e c^3} > 4\pi,
\end{equation}
where $\sigma_T$ is the Thomson cross section.
The only caveat attaching to this statement is an order unity correction
dependent on the details of the spectrum.  In virtually all work since, it has been assumed that when $l \gg 1$, few
high energy photons can escape the source region, although
subsequent pair annihilation can partially restore them ({\it e.g.},
Guilbert, Fabian \& Rees 1983; Svensson 1987; Blandford 1990).  This argument
has been used
in many contexts because sources rich in $\gamma$-rays are very often
copious sources of softer photons as well, and are also often variable enough
to indicate causality bounds on the source size which result in large estimated
compactnesses.

    However, it is not always true that $l \gg 1$ implies
$\tau_{\gamma\gamma} \gg 1$.  In
particular, when the sources of high and low energy photons are geometrically
separated, the very pair production which has been thought to prevent
$\gamma$-ray escape can instead ensure it.  Because high compactness
generally also entails a significant Compton optical depth (Guilbert {\it et
al.} 1983), an optically thick Compton scattering layer can form between
the two source regions.  High energy photons incident upon this surface
from one side
are either absorbed in pair production reactions, or else lose much of their
energy by Compton recoil; low energy photons, which strike it from the
other side, are scattered
away, so that only a few cross the surface onto the side where the
$\gamma$-ray source is located.  The net result is that those high energy
photons directed away from the surface can escape freely, while the energy of
the $\gamma$-rays directed into the surface is transformed into a relativistic
outflow of mixed $e^{\pm}$ pairs and photons.

    The remainder of this paper is devoted to working out a ``toy-model"
version of this idea so as to illustrate its qualitative features.  It
would be premature at this stage to use these ideas as the basis for a
detailed model of any
particular source or class of sources; our goal instead is to work out
a number of gross features and scaling properties of this class of model
to provide guidance for future, more realistic applications.  Hence the
following description is necessarily highly idealized.

\section{Problem Definition}

     If the original photon sources radiate isotropically, there is no
spherically symmetric system in which this screening mechanism may operate.
The problem is that the high energy photons
must eventually pass through the low energy photon source region
in order to escape.  When they do so, they see an isotropic distribution
of pair production partners, and the optical depth will be correctly
approximated, at least to order of magnitude, by the usual scaling from
the compactness.

     However, there are a number of only slightly more complicated (and
possibly even physically plausible) geometries which do permit operation
of this mechanism.  In this paper
we will analyze one of them in a fashion that is deliberately as simple and idealized as possible.  Real sources are undoubtedly
more complicated.  Our goal here is to highlight the basic physical
principles underlying the scheme.

With accreting black hole systems (whether stellar binaries or active galactic
nuclei) in mind, we choose a source geometry in which
the high energy photons are generated isotropically with luminosity $L_\gamma$
in a small spherical source of radius $s$, while
the low energy photons (``X-rays") are radiated isotropically with
luminosity $L_x$ from a ring of radius $a$
whose center is the $\gamma$-ray source.
In real sources, we expect that the regions of X-ray and $\gamma$-ray emission 
are likely to overlap.  Nonetheless, internal gradients in the relative hardness
of the spectrum are also quite likely.  Our picture of complete separation
between the $\gamma$-ray and X-ray sources is intended as an idealization
which allows the basic idea to be seen more clearly through ease of
calculation.  In keeping with our effort to present the simplest possible
case, we suppose that the
spectra emitted by the small sphere and the ring have the same shape, but do
not overlap in energy.  That is, the sphere emits a spectrum (in erg/erg)
$\propto x^{-\alpha}$ for $1 < x_l \leq x \leq  x_{max}$, while the spectrum
of the ring is likewise $\propto x^{-\alpha}$, but for $x_{min} < x \leq 1$.
Here and throughout the paper we give photon energies in units of $m_e c^2$.

  This separation of high energy photons from low does not automatically
guarantee that the $\gamma$-rays can escape from their source because
the pair production cross section for pairs of photons well above threshold
while small, is not zero.  In order to avoid excessive pair production
within the $\gamma$-ray source, the parameters must satisfy the condition
\begin{equation}
x_l > \left[ \left( {1 -\alpha \over 1 + \alpha}\right)
\left({l_\gamma \over 4\pi}\right){0.2 \over x_{max}^{1-\alpha} -
x_l^{1-\alpha}}\right]^{1/(2 + \alpha)}.
\end{equation}
Here, and elsewhere in the paper, when we need to integrate over the
the $\gamma - \gamma$ cross section, we approximate its energy dependence
by
\begin{equation}
\sigma_{\gamma\gamma} \simeq 2g \sigma_T (x_1 x_2)^{-1},
\end{equation}
where $x_1$ and $x_2$ are the energies of the two photons participating,
and $g \simeq 0.2$.  This is an approximation to the form given by
Gould \& Schreder (1967), corrected by Brown, Mikaelian \& Gould (1973).
When $\alpha = 1$ (which is the center of the spectral distribution for
the AGN detected by EGRET: Hartman \etal 1994), the ratio
$(1-\alpha) /(x_{max}^{1-\alpha} - x_l^{1-\alpha})$ becomes
$1/\ln (x_{max}/x_l)$,
so that $x_l$ must be at least $\sim (l_\gamma/4\pi)^{1/3}$.

   The question naturally arises whether such a hard spectrum could
be created.  Here we merely point out, in a very
schematic way, a few possible mechanisms.  In principle, a thermal
source with $kT \gg m_e c^2$ qualifies, though it may be difficult in
practise to sustain such a source.  Perhaps a more plausible mechanism
would be for the $\gamma$-rays to be
produced by inverse Compton scattering.  If the seed photons have
energy $x_o < x_l^{-1}$,  and the electron distribution function
is a power-law between $\gamma_{min} = (x_l/x_o)^{1/2}$ and $\gamma_{max} =
x_o^{-1}$, the resulting spectrum runs from $\sim x_l$ to $x_{max} = \gamma_{max} = x_o^{-1}$, and there are
{\it no} photons in the seed population that
may pair produce with any of the $\gamma$-rays.  Interestingly, electron distribution functions with $\gamma_{min} \sim 100$ have also been suggested to exist in relativistic AGN jets in order to avoid
pair annihilation constraints (Ghisellini \etal 1992) or to
avoid Faraday depolarization (Wardle 1977).

   In fact, the lower bound on the energies of photons produced in the
$\gamma$-ray source can
be softened somewhat when one takes into account a nonlinear transfer effect.
Ordinarily $\gamma$-ray transfer is considered in the linear approximation;
that is, it is assumed that the pair production opacity is fixed.
However, when the number density of X-rays is not much greater than the
number density of $\gamma$-rays, that assumption is not justified, and
the transfer problem becomes nonlinear because the result of absorbing
$\gamma$-rays in pair production events is to decrease the X-ray density
as well, thereby diminishing the pair production opacity.  To estimate
the magnitude of this effect, we make the approximation that the
greatest impact on the density of X-rays of energy $x$ is due to
pair production with $\gamma$-rays of energy $\epsilon = 1/x$, \ie those
right at threshold.  If that is the case, the nonlinear response of the
opacity permits $\gamma$-rays to escape---despite apparently large
compactness---when $n_x < \epsilon^2 n_\epsilon$, where $n_x$ is
the number density per unit energy of the X-rays at dimensionless
energy $x$ and $n_\epsilon$ is the
number density per unit energy of the $\gamma$-rays.  This condition
is equivalent to requiring a spectral index $\alpha < 0$
between $x$ and $\epsilon$.

 In at least two sources (CTA 102 and 3C 454.3: Blom
\etal 1995), the observed spectra actually come close to meeting this
criterion.  In both, the observed spectral index is $ < 0.3$ from
$\sim 1$ keV to $\sim 10$ MeV.
Because there are no observations of these objects between 2 keV and 1 MeV, and
the measurements below 20 MeV are only upper bounds, the spectrum could be even
harder than this in the low-energy $\gamma$-ray band.  In addition,
since observed spectra integrate over the entire source, local regions
within these sources
could produce still harder spectra.  Thus, sources such as these may
be examples of places where there are indeed fewer low-energy
pair-production partners than $\gamma$-rays.

\section{Formation and Location of the Screening Surface}

    Suppose that both the X-ray source and the $\gamma$-ray
source turn on gradually.  The
two classes of photons are initially well-mixed everywhere.  As the
photon densities increase, the pair production rate also increases.

   When first created, the pairs move with the net momentum of the photons
which produced them (the sum is dominated by the high energy photon, of
course). From that point on, their motion is subject to Compton
scatters against the photons and collisions with other pairs.
Because these particles are relativistic,
gravity is negligible unless these events take place close to
the event horizon around a black hole.  The X-rays see a
significant Compton opacity (as a result of the pair production), while the
$\gamma$-rays see both Compton opacity (in the Klein-Nishina regime) and
pair-production opacity (which is greatest for
the highest energy photons).  Consequently, the local radiation force
depends significantly on the optical depth from the point of observation to
the various photon sources.  Pairs close to the $\gamma$-ray source will
tend to be pushed outward, while pairs just inside the X-ray ring will be pushed
inward.

    The pair densities in the zones closest to the two sources (just outside the
origin, and just inside the ring) are further reduced by an additional optical
depth effect: because the $\gamma$-rays and X-rays are no longer mingled,
the pair production rate falls.  Suppose we define these partially-evacuated
zones by the surfaces on which the total optical depth from the nearest photon
source is unity.  Then the radiation forces identified in the previous paragraph
lead to a continuous expansion of their volume.

     Eventually, the bounding surfaces of the evacuated zones must meet.  These
merged surfaces are the screening surface.  The electrons within this
screening surface absorb momentum from the radiation by two mechanisms: by Compton scattering photons incident from either side, and by pair production
inside the surface wherever the high and low energy photons can mix.  If
the electrons flow relativistically, the
sign (relative to the direction of the electron velocity) of the net momentum exchange due to Compton scattering external
photons can change from positive to negative. Momentum
is also carried out of the surface by photons created within it which escape.
The surface ceases to move when the net momentum transferred to it is
directed entirely within its tangent plane, \ie when the normal components
of the momentum absorbed on either side exactly balance.

    To find the location of this equilibrium surface, consider the limit in
which the normal component of the incident radiation force is
proportional to the normal
component of the radiation flux.  This limit is achieved when the albedo
of the surface is small and it is optically thick.

    Call the axis of the ring the $z$-axis.  The
screening surface is then defined by $z_s(r)$, where $r$ is the cylindrical
radius.  By azimuthal symmetry, we need not consider azimuthal components
of the radiation force.  In the limit that the screening surface is
very thin compared to the ring radius $a$, the condition that the net
normal component be zero may be expressed as
\begin{equation}
{L_{\gamma} \over L_x} {|\hat R \cdot \hat n|^2 \over r^2 + z_s^2} =
\int_{-\Delta \phi}^{+\Delta\phi} \, {d\phi \over 2\pi} \, {|\hat R_x (\phi)
\cdot \hat n|^2 \over d^2 (\phi)},
\end{equation}
where $\hat R$ is the unit vector in the spherical radial direction,
$\hat n = [-(dz_s/dr) \hat r + \hat z]/[1 + (dz_s/dr)^2]^{1/2}$ is
the unit vector normal to the surface,
$\hat R_x (\phi) = [(r - a \cos\phi)\hat r - a\sin\phi \hat y +
z_s \hat z]/d(\phi)$,
and $d(\phi) = [r^2 - 2ar\cos\phi + a^2 + z_s^2]^{1/2}$.  The factors
$|\hat R \cdot \hat n|^2$ and $|\hat R_x \cdot \hat n|^2$ appear
because the pressure is proportional to the rate at which photons
cross the surface times the component of their momentum parallel to
the surface normal.  The half-opening
angle of the portion of the ring from which X-rays may arrive at
$(r,z_s)$ without being
blocked by Compton opacity is $\Delta\phi$.  The actual value of
$\Delta \phi$ as a function of $z$ is hard to determine {\it a priori}
because it depends on the diameter of the screening surface at lower
altitudes.  We adopt the simplistic approximation that the blocking
is what would occur if the screening surface closer to the equatorial
plane were simply a cylinder of radius $a$ for $r > a$, or a cylinder
of radius $r$ for $r < a$.  That is, $\Delta\phi = \cos^{-1} (a/r)$
for $r > a$, or $\Delta\phi = \cos^{-1}(r/a)$ for $r < a$. 

   Equation 4 implicitly defines
an ordinary differential equation for $z_s (r)$ parameterized by
$L_{\gamma}/
L_x$.  The appropriate initial condition for this differential equation is
$r(z_s = 0)$ such that the radial component of the net force is identically
zero.  Solving equation 4 yields ``hour-glass" shapes which widen into cones
at large distance.  The opening angles of these cones increase
with increasing $L_\gamma/L_x$ (Fig. 1).

   In reality, the limit of large optical depth and zero albedo (or
at least equal albedos for $\gamma$-rays and X-rays) is
achieved at best imperfectly, but the degree of departure
from this limit may be used to calculate
an ``effective" $L_{\gamma}/L_x$. In \S 5 we show that essentially all
the $\gamma$-rays incident on the screening surface are absorbed; after
calculating the bulk Lorentz factor of the flow (\S 6.1) and the characteristic
temperature associated with the majority population of pairs
within the surface (\S 6.2), one
may compute the effective albedo to X-rays, and therefore the enhancement
to the absorbed X-ray momentum flux.

\section{How the $\gamma$-rays Escape: The Reduction of $\tau_{\gamma\gamma}$
on the Axis}

If the photons all simply streamed freely away from their sources, the
density of X-rays originating from angle $\phi$ on the ring
would fall $\propto d^{-2}(\phi)$.
The optical depth to pair production for a $\gamma$-ray of energy $x$
traversing the region would then be
\begin{equation}
\tau_{\gamma\gamma}^{(o)}(x) \simeq g {1 -\alpha \over 1 + \alpha}
{(x/2)^{\alpha} \over 1 - x_{min}^{1 - \alpha}} {l_x \over 4\pi}.
\end{equation}
The expression we
show assumes that the photon direction is parallel to the $z$-axis, but
there is only a weak dependence on polar angle in the sense that the
optical depth
is smallest on the axis and greatest in the equatorial plane.  Most
of the optical depth is accumulated inside $R \sim a$; at greater distances,
the optical depth from $R$ to $\infty$ scales as $(R/a)^{-2\alpha -3}$.
This scaling results from the effective increase in the pair production
threshold as the photon trajectories become more and more parallel.
Thus, if no other process happens first, virtually all emitted $\gamma$-rays
would be transformed into pairs when $l_x \gg 1$.

   However, the presence of intervening, relativistically streaming, electrons
and positrons alters the density, energy spectrum, and direction of the
X-rays.  Call the
bulk speed parallel to the surface $\beta c$, and the bulk Lorentz factor
$\Gamma$ (see \S 6.1 for estimates of their likely magnitude).  We will
label all quantities measured in the moving frame by primes.

   When the Thomson depth $\tau_s^{\prime}$ of the screen (measured
along the normal in the moving frame; this is not a Lorentz invariant
because the direction of the apparent normal is not invariant) is more than a
few, the photon number flux penetrating through the screen is reduced
substantially.  Scattering alone reduces their flux by $\sim 1/\tau_s^{\prime}$.
In addition, in the moving frame of the pair fluid, photons take on average
a time $\tau_s^{\prime}h/c$ to diffuse through the screening surface.
Here $h$ is the transverse thickness of the screen.  Photons which enter the
screen at lab frame coordinate $\eta$ (distance along the screen
measured outward from the equatorial plane) then emerge at $\eta + \beta \Gamma \tau_s^{\prime}h$.  By the time they leave the surface, their density
has been reduced by a further factor because the circumference of the
screening surface has increased.  If the surface is nearly conical
(as Fig. 1 shows that it often is), the photon density is reduced by
the factor
$1/[1 + \beta \Gamma \tau_s^{\prime} h (d\ln r/d\eta)]$.  Moreover, their
direction of motion has been changed so that (in the lab frame) their directions
are concentrated within an angle $\Gamma^{-1}$ of the local surface
tangent.  Some reflected photons may strike the surface multiple times
(see Fig. 1);
the importance of this effect increases with $\Gamma$.  Note also that when $\beta\Gamma\tau_s^{\prime}h > \eta$, the optical
depth of a fluid element changes significantly from the beginning of this
process to the end, so a suitably averaged value of $\tau_s^{\prime}$ should
be used in a more careful calculation of this effect.

   Compton scattering inside the surface causes the X-rays to lose
some energy by recoil, but on balance to
gain energy.  Most of the electrons in the surface thermalize before
annihilating (Lightman and Zdziarski 1987); their temperature in
the bulk frame $T^{\prime}$ (measured in units of $m_e c^2$) should be more
than $ T_C$, the Compton temperature of the incident X-rays, because they
are also exposed to the much harder photons generated by the pair cascade
inside the surface (\S 6.2).  The (non-relativistic) Compton-$y$ parameter
${\tau_s^{\prime}}^2 T^{\prime}$ will then very likely be
large enough to substantially alter the X-ray spectrum inside the
screening surface (see \S 5).

   The net result of all these effects--the reduction in flux by scattering,
the advection to larger distance, the redirection more nearly parallel to the
surface, and the increase in mean photon energy--is to drastically reduce
$\tau_{\gamma\gamma}$ on the axis.  Because most of $\tau_{\gamma\gamma}^{(o)}$
is due to X-rays found within $R \sim a$ of the origin, if the screening
surface is Compton thick to at least $z \sim a$, it can effectively
shield the $\gamma$-rays from the greatest part of the pair production
opacity that would otherwise exist.

\section{Pair Balance and the Compton Depth of the Critical Surface}

   Our next task, therefore, is to estimate the Compton depth of
the screening surface as a function of position.  Where the surface is
optically thick to Compton scattering, it effectively shields those
$\gamma$-rays whose paths are close enough to the axis that they miss
the surface; in addition, those X-rays striking the surface where
it is Compton thick are reflected outward, contributing to the X-ray
flux seen in directions nearly tangent to the surface.

\subsection{Pair Yield}

   To make this estimate, we must find the rate at which the energy of
$\gamma$-rays striking the surface is transformed into pairs.  The problem treated here is similar in many respects to the pair
balance calculations of Svensson (1987) and Zdziarski (1988).
In both our case and theirs, a pair cascade
is initiated by high energy photons reacting with softer photons, and
the subsequent generations of pairs lose energy primarily by inverse
Compton scattering.  Likewise, in all cases the pairs are trapped in the region
(by magnetic fields? by scattering against plasma waves?) while photons
can escape.

   However, there are also a number of points of
difference.  Some may be accounted for by appropriate adjustments.
In both Svensson's and Zdziarski's calculations, all the energy injected
into the region
eventually found its way into the pair cascade; in our case, it is
possible for some of the photons to cross the region without being
absorbed (although in the cases of greatest interest, we expect the
$\gamma-\gamma$ optical depth to be large over the interesting range
of photon energies). Their calculations
assumed isotropic photons, while in our problem they are directed;
this difference can be removed by appropriate renormalization of
energies and compactnesses. 

   There are also a number of more significant contrasts, however.
In their problems, the length scales governing the
crossing time and the optical depth were the same as the length scale in
the compactness; here they are different in the sense that both the
crossing time and the optical depth are smaller in our context
by the ratio $h/a$.
In their case, the soft photons injected into the region were
Comptonized by the full optical depth; in this case, they are Comptonized
by reflection, so that most photons experience fewer scatterings than
would be the case when they are emitted throughout the volume under
consideration.  The
final significant contrast is that in Svensson's calculation all Compton
scattering takes place in the Thomson limit, while in Zdziarski's
calculation the most important Compton scattering events are in the
Klein-Nishina regime; in our case, either limit may be appropriate,
depending on the spectra of the two sources. 

   On the basis of these differences, we expect that the pair yield
$Y$, the ratio between the total particle production rate and the
absorbed $\gamma$-ray energy in units of $m_e c^2$,
is likely to be somewhat smaller in our case than in theirs. This means
that when $l_\gamma \gg 1$, $Y$ is at most $\sim 0.1 L_T/L_\gamma$,
where $L_T$ is that portion of the absorbed $\gamma$-ray luminosity
which is deposited in pairs which scatter predominantly in the
Thomson limit.

\subsection{Fraction of $\gamma$-rays absorbed}

   To find the portion of $L_\gamma$ striking the surface which is absorbed,
we must re-estimate
the $\gamma-\gamma$ optical depth along those directions including the effects
of the screening surface.
Because the energies and directions of the X-rays reflecting off the surface
are quite different from what they were initially, the $\gamma -\gamma$
optical depth cannot be directly computed on the basis of the X-ray
intensity distribution radiated by the ring.  We therefore divide the
problem into
two parts: the $\gamma-\gamma$ optical depth due to never-scattered
X-rays, and the $\gamma-\gamma$ optical depth due to scattered X-rays.

   Within
the surface itself, never-scattered X-rays can be found in significant numbers
only in an outer layer roughly one Compton depth thick.  Therefore, most
of the pair production optical depth due to never-scattered photons is 
found on the portion of the ray beyond the surface.  For rays
encountering the surface at $R \sim a$, the contribution these photons make to
$\tau_{\gamma\gamma}$ is smaller than $\tau_{\gamma\gamma}^{(o)}$ by
a factor of a few; the reduction factor for rays striking the surface at larger
distance is $\sim (a/R)^{2\alpha + 3}$ for the same reason that the
outer contributions to $\tau_{\gamma\gamma}^{(o)}$ scale this way.

   Those X-rays reflected by Compton scattering are changed both in energy
and direction.  If $T^{\prime} < 1$, the results of Lightman \& Rybicki (1980)
give at least an approximate description of the reflected spectrum in
the moving frame.  When the Compton $y$-parameter is larger than $\simeq \ln
(T^{\prime}/x_{min}^{\prime})$, a fraction $f_W = \max\{1,[(4T^{\prime}/
\ln(T^{\prime}/x_{min}^{\prime})]^{1/2}\}$ of the photons are
scattered into a Wien distribution at the
temperature $T^{\prime}$, with the remainder distributed into a power
law with energy spectral index $\simeq 0$.  Note that this fraction is
independent of $y$ (for values greater than the minimum) because we are
interested in the reflected spectrum, not
the spectrum of photons created inside the plasma.  In the lab frame, the photon
energies are increased by $\Gamma (1 + \beta)$, and, to the extent
that $\Gamma$ is greater than unity, they are beamed parallel to
$\hat t$, the unit vector tangent to the surface.

    Because the pairs in the screening surface move relatively slowly
when $R < a$ (see \S 6.1), relativistic boosting and beaming are not
very important in that region.  To estimate the pair production
optical depth due to the scattered photons, we make the crude approximation that
in this region the density of the photons is roughly what it would
have been without reflection, and that their angular distribution is
broad enough that few scattered X-rays move parallel to the $\gamma$-rays.
The optical depth is then due to two contributions: that due to the
flat power-law segment of the scattered photon spectrum, and that due
to the Wien segment.  Integrating these spectral shapes over the
energy-dependent pair production cross section gives
\begin{equation}
{\tau_{\gamma\gamma}^{(scatt)} \over \tau_{\gamma\gamma}^{(o)}} \simeq
{1 - f_W \over x_{min}(x/2)^{\alpha}\ln(4T^{\prime}/x_{min})} +
{3 f_W \over x_{min} x^{1+\alpha} T^{\prime}},
\end{equation}
where the expression for the Wien contribution is valid for $x > 2/T^{\prime}$.
Not surprisingly, the result of Compton upscattering is to increase
the opacity for lower energy $\gamma$-rays and decrease the opacity for
higher energy $\gamma$-rays.

   Farther out along the surface, relativistic effects can be expected
to be more significant (though $\Gamma$ is unlikely to ever be $\gg 1$,
as demonstrated in \S 6.1).  Because the direction of the surface tangent
tends to a constant at large distances (Fig. 1), relativistic beaming
will have the effect of reducing the angle between the $\gamma$-rays and
X-rays, thereby diminishing the opacity.

   Two other effects may also contribute to $\gamma - \gamma$ opacity, but
their magnitude is harder to estimate.  Bremsstrahlung by the pairs
may increase the number of low energy photons, particularly in the side
of the screen nearer the $\gamma$-ray source.  While the total number
of photons entering the screen is always likely to be dominated by the
external X-ray source (unless $L_\gamma /L_x \gg 1$), most of the externally
created photons that penetrate far into the screen are upscattered to high
enough energies that their pair production opacity is reduced.  Synchrotron
radiation may also contribute significant numbers of soft photons.  Because
high energy pairs are continually created by the absorption of high energy
$\gamma$-rays, the steady-state electron distribution function will contain
a nonthermal tail at high energies.  However, the amplitude of this tail
is very difficult to determine because of the many possible cooling mechanisms
for these electrons (\eg inverse Compton scattering, synchrotron radiation,
Coulomb collisions with other electrons and positrons, plasma wave scattering).

   Summarizing these arguments, we expect that if $l_x/4\pi > 1$,
the fraction of the $\gamma$-ray energy striking the surface which is
absorbed remains close to unity out to distances a few times $a$.

\subsection{Equilibrium Compton optical depth}

    We may now estimate the optical depth as a function of position by balancing annihilation against pair production ({\it cf.} Guilbert {\it et al.} 1983).
In the frame of the streaming pairs, the rate of pair production is
given by the Lorentz-transformed rate at which $\gamma$-ray flux is
absorbed times the pair yield $Y$.  Note that the relevant flux comprises
only those photons above the pair production threshold.
In terms of the Compton depth of the
surface, the positron density in the moving frame may be written as
$n_+^{\prime} = \tau_s^{\prime}/(2\sigma_T h)$.  Thus, we find
\begin{equation}
\tau_s^{\prime} = \left\{ {Y \over
\Phi(T^{\prime})} {1 - \alpha \over x_{max}^{1-\alpha} - x_l^{1-\alpha}}
{f_{\gamma} l_\gamma \over 4\pi} \left({ha \over R^2} \right)^2 
\left[\Gamma (1 - \vec\beta \cdot \hat R)\right]^3 |\hat R \cdot \hat n|
\int_{\max(x_l, 1/[\Gamma(1 - \vec\beta \cdot\hat R)])} \, dx \, x^{-\alpha}
\right\}^{1/2},
\end{equation}
where $\Phi(T) \simeq 3/8$ is the pair annihilation rate at
temperature $T$ in
units of $\sigma_T c$, and $f_{\gamma}$ is the fraction of the $\gamma$-ray
energy absorbed.
In the equatorial plane, where $\Gamma \simeq 1$, we expect $\tau_s^{\prime}
\simeq [Y f_{\gamma}l_{\gamma}/(4\pi)]^{1/2}(h/R)$.  Equation 7 shows,
however, that the optical depth should typically fall rapidly
for $\eta > a$: as the pairs stream out, their density falls;
at large distances the photon directions become nearly parallel, so that
the pair creation threshhold rises and the primary production rate falls;
and the photon directions become nearly tangent to the screening surface, so
that the rate at which photons enter the surface falls.  In other words,
when $l_x/4\pi > 1$ and $l_{\gamma}/4\pi \gg 1$, the screening surface is
optically thick to Compton scattering out to distances at least $\sim a$.
It is in these circumstances that the surface is able to effectively
protect the $\gamma$-rays moving within the open cone from pair
production with the X-rays.

\section{Bulk Properties of the Surface}

\subsection{Flow Speed}

    The momentum of the flow along the surface is carried both by the
pairs and by the photons trapped within the surface.  New momentum is
added by the arrival of both $\gamma$-rays and X-rays; to the extent
that X-rays are reflected, particularly with additional energy,
they carry away momentum.  Photons produced
by annihilation or bremsstrahlung can remove momentum as they escape from
the flow.  At the same time, the inertial density of the surface is increased
by energy absorbed from the incident X-rays and $\gamma$-rays, and diminished
by escaping photons.  In the steady state and assuming pure absorption (and
no intrinsic radiation), the conservation equations for
momentum along the surface and the inertial density combine to form the pair
\begin{equation}
{4\pi c^3 \beta \rho h \Gamma \over [1 + (dz_s/dr)^2]^{1/2}}
{\partial \beta \over
\partial r} = {f_{\gamma}L_{\gamma} \over r^2 + z_s^2}
|\hat n \cdot \hat R| \left[ |\hat t \cdot \hat R| - \beta\right] +
\int_{-\Delta\phi}^{+\Delta\phi} \, {d\phi \over 2\pi} L_x {|\hat n \cdot
\hat R_x (\phi)| \over d^2 (\phi)} \left[|\hat t \cdot \hat R_x (\phi)| - \beta
\right]
\end{equation}
and
\begin{equation}
{4\pi c^3 \beta \over [1 + (dz_s/dr)^2]^{1/2}}{\partial (\rho h \Gamma) \over
\partial r} = {f_{\gamma}L_{\gamma} \over r^2 + z_s^2}
|\hat n \cdot \hat R|  +
\int_{-\Delta\phi}^{+\Delta\phi} \, {d\phi \over 2\pi} L_x {|\hat n \cdot
\hat R_x (\phi)| \over d^2 (\phi)}.
\end{equation}
Derivatives with respect to distance along the surface are related to
derivatives with respect to $r$ by the factor $[1 + (dz_s/dr)^2]^{1/2}$.
The factors proportional to $|\hat n \cdot \hat R|$ and
$|\hat n \cdot \hat R_x|$ are the geometrical corrections for $\gamma$-ray
and X-ray flux, respectively, entering the surface.  The dot products
of $\hat R$ and $\hat R_x$ with $\hat t$ are the geometrical corrections
accounting for the portion of the incident momentum (from $\gamma$-rays
and X-rays, respectively) parallel to the surface.  The subtracted
factor of $\beta$ is required because the energy absorbed along with
momentum increases the inertia density, so that the speed does not increase
unless the geometric factor exceeds $\beta$.

    Solutions of these equations are specified by two initial conditions,
one on $\beta$ and the other on $\rho h \Gamma$.  At $r(z_s = 0)$, it is
obvious by symmetry that $\beta$ must be zero (for numerical solutions
we set it to 0.01 in order to avoid artificial divergences).   In appropriate
dimensionless units, the inertia density $\rho h \Gamma = (\mu_e/m_e)\tau_T/l_x$, where $\mu_e$ is the inertia per electron.  $\mu_e$ can
be greater than $m_e$ due both to the relativistic motions of the electrons
and the effective inertia of the radiation trapped in the surface.
Again to avoid numerical problems, we set this initial condition to a small
fraction of unity; in practise, it rises to its asymptotic value so swiftly
that there is little dependence on exactly how we choose this
initial condition provided it is small.

     Numerical solution of equations 4, 8, and 9 demonstrates that solutions
extending to infinity exist for all $L_\gamma/L_x > 0.08$.  When this ratio
is smaller, the surface closes over the origin, and $\gamma$-rays cannot
escape freely in any direction.  We expect that in this case the ultimate
structure would be a quasi-cylindrical pair-rich wind, but we have not
investigated it in detail.

    For $L_\gamma/L_x > 0.08$, the surface
becomes asymptotically conical with an opening angle that increases with
increasing $L_\gamma/L_x$ (Fig. 1).  When $L_\gamma/L_x > 10$, the half
opening angle
is nearly $\pi/2$, \ie the only region from which X-rays are not excluded
is a thin wedge in the equatorial plane.  However, because a careful
calculation of the pair balance is beyond the scope of this paper, it
is possible that further collimation toward the polar direction occurs
in the region $a < r < few \times a$.  When the screening
surface becomes optically thin (at $r \sim a$?), the complete
absorption approximation is
no longer appropriate.  Instead, the radiation force is simply proportional
to the flux times the opacity.  Because Klein-Nishina effects reduce the
opacity of the $\gamma$-rays, the effect of the X-rays is comparatively
enhanced.  Provided $r$ is not too much larger than $a$, that translates
to an increase in the component of the radiation force parallel to $\hat z$.

    Both the tangential speed $\beta$ and the inertia density $(\mu_e/m_e)
\tau_T/l_x$
quickly reach asymptotic values (Figs. 2 and 3).
At first glance it might seem surprising that the asymptotic speed is
only mildly relativistic unless $L_\gamma \gg L_x$.  After all, when a
high energy $\gamma$-ray combines
with an X-ray to create a pair, the electron and positron both have very
large Lorentz factors.  However, in this model most of the pairs are
created where $z_s/a < 1$, and the momenta of the $\gamma$-rays and
X-rays are almost oppositely directed.  Consequently, the plasma begins
with a fairly large ratio of inertia density to momentum, and by the
time the $\gamma$-rays are more nearly parallel to the flow, little can
be done to accelerate it.  It is possible that greater speed might be
found with allowance for intrinsic radiation. However, the same radiation which keeps the inertia density small also removes momentum from
the flow. We have modelled this effect by a phenomenological non-zero albedo,
and confirmed that $\beta$ depends only weakly on the size of the albedo (in
fact, increasing the albedo to 0.5 actually {\it decreases} the
asymptotic $\beta$ slightly).
On the other hand, if for some reason the albedo decreased with distance,
the asymptotic speed might be increased.

    The asymptotic inertia density quickly reaches a constant value because
the rate of energy absorption by the surface decreases rapidly with increasing
distance.  The flux striking the surface falls so quickly both because of
increasing distance and because the surface becomes nearly tangential
to the photon paths it intercepts.  $(\mu_e/m_e)\tau_T/l_x$ is a very slowly
increasing function of $L_\gamma/L_x$; for $L_\gamma/L_x \sim 1$,
it is $\simeq 0.3$.  Provided $\mu_e/m_e$ is not too large, the surface
is indeed Compton thick whenever $l_x$ is significantly greater than unity.
It should be borne in mind, however, that our model equations do not include
the effect of photon leakage out of the surface as its optical depth
decreases at large distance.  Consequently, these asymptotic values of
the inertia density are likely to be overestimates at $(r^2 + z_s^2)^{1/2} \gg
a$.  At larger distances, the flow speed could well be larger than our
model predicts because the pairs are not forced to share their energy with
so many trapped photons.

\subsection{Pair Temperature}

    The temperature of those pairs which have cooled into a thermal
distribution is determined primarily by Comptonization balance.  Generically,
we expect bremsstrahlung to play a secondary role in cooling: relative
to inverse Compton scattering of the incoming X-ray photons, the
bremsstrahlung luminosity when $T^{\prime} \sim 1$ is $\sim \alpha_{fs}
(\tau_s^{\prime}/l_x^{\prime})(a/h)$, where $\alpha_{fs}$ is the fine structure
constant and $l_x^{\prime}$ is the X-ray compactness on the scale $a$
as viewed from the frame of the screening surface.  The local photon
spectrum participating in this Comptonization equilibrium combines incident
$\gamma$-rays which have avoided pair production, incident X-rays 
which have penetrated into the surface with little energy change, photons
originally in the X-ray power-law
which have scattered sufficiently against the electrons to enter a Wien
distribution, and the smaller number of photons which have undergone
large increases in energy by Compton scattering against nonthermal
electrons.  Because the effectiveness of scattering declines for those
photons which interact in the Klein-Nishina regime, the most likely
temperature is a fraction of unity, that is, $\sim 100$ keV.

\subsection{Geometrical Thickness}

    When $\tau_s^{\prime} > 1$, the annihilation time for an electron
is shorter than the sound wave crossing time.  The thickness of the
surface is therefore controlled more by pair balance (that is, radiation
transfer) than by pressure balance.  Suppose, then, that $x_*^{\prime}$
is the $\gamma$-ray energy in the bulk frame at which $Y(x^{\prime})
L_\gamma^{\prime} (x^{\prime})$ is greatest.  The pair balance as a function of
distance $\xi$ into the surface is then determined by two coupled
differential equations which describe the radiation transfer:
\begin{equation}
\mu^{\prime} {dn_\gamma^{\prime} \over d\tau^{\prime}} =
-\left({\zeta\Phi n_{xo}^{\prime} \over 2 x_*^{\prime} Y}\right)^{1/2}
\left({\tau^{\prime}\over\tau_s^{\prime}}\right)^{1/2} {n_{\gamma}^{\prime}}^{1/2}
\end{equation}
and
\begin{equation}
{d\tau^{\prime}\over d\xi} = 2 n_+^{\prime}\sigma_T
\end{equation}
In these equations, $\mu^{\prime} = |\hat n^{\prime} \cdot \hat R^{\prime}|$,
$n_{\gamma}^{\prime}$ is the number density in the
bulk frame of photons of energy $x_*^{\prime}$, $n_+^{\prime}$ is the
density of positrons, and $\zeta$ is the pair production cross section
at $x_*^{\prime}$ in units of $\sigma_T$ averaged over the X-ray
spectrum.   The first equation describes the transfer of $\gamma$-rays
subject to absorption by pair production with X-rays.  The second equation
defines the Compton optical depth scale relative to the distance scale.  The
Compton opacity is proportional to
${n_{\gamma}^{\prime}}^{1/2}$ because the pair density (assuming pair
balance) is proportional
to the square root of the $\gamma$-ray density:
\begin{equation}
n_+^{\prime} = \left({\zeta n_{\gamma}^{\prime} n_{xo}^{\prime} \tau^{\prime}
 x_{*}^{\prime} Y \over 2\tau_s^{\prime}\Phi}\right)^{1/2}
\end{equation}
Here we have also
made the approximation that Compton scattering opacity dominates
pair production opacity for X-rays, so that the X-ray number
density decreases linearly with Thomson depth $\tau^{\prime}$ from $n_{xo}^{\prime}$
at the edge where they strike the surface, to zero at the edge where the
$\gamma$-rays enter.  However, the mean energy of the X-rays changes as a
result of Compton scattering; this variation
is absorbed into $\zeta$.

  Equation 10 is readily solved in terms of $\tau^{\prime}$:
\begin{equation}
n_{\gamma}^{\prime} = \left[ {n_{\gamma o}^{\prime}}^{1/2} -
{{\tau^{\prime}}^{3/2} \over 3 {\tau_s^{\prime}}^{1/2} \mu^{\prime}}
\left({\zeta\Phi n_{xo}^{\prime} \over 2 x_{*}^{\prime} Y}\right)^{1/2} \right]^2,
\end{equation}
where $n_{\gamma o}^{\prime}$ is the number density of $\gamma$-rays at
the $\tau = 0$ edge. 
Substituting this result in the opacity differential equation, two
characteristic length scales are revealed:
\begin{equation}
\lambda_1 = {1 \over \sigma_T} \left({\Phi \tau_s^{\prime} \over 2 \zeta
x_*^{\prime} Y n_{xo}^{\prime} n_{\gamma o}^{\prime}}\right)^{1/2}
\end{equation}
and
\begin{equation}
\lambda_2 = {6 \mu^{\prime}\tau_s^{\prime}  \over \zeta n_{xo}^{\prime}\sigma_T}.
\end{equation}
The first of these describes the characteristic thickness of the surface's
edge on the side facing the $\gamma$-ray source; the second describes
the thickness on the side facing the X-ray source.  Typically we expect
$\lambda_1$ to be rather greater than $\lambda_2$ because $n_{xo} \gg
n_{\gamma o}$.  The total thickness is, of course, of order their sum.

    Because both scales depend on $\tau_s^{\prime}$, which, in turn,
depends on the sum of the two lengthscales, it is possible to use these
relations to solve explicitly for $h/a$.  This procedure is simplest
at $z_s = 0$ where $\beta = 0$, so the relativistic transformations
are null.  When, in addition, $1 - r_s/a \ll 1$ (a condition which
applies whenever $L_\gamma/L_x$ is not too small), the incident X-ray
density is
\begin{equation}
n_{xo} \simeq {l_x \over a \sigma_T [1 - (r_s/a)^2]}.
\end{equation}
Using this X-ray density, we find the optical depth at $z_s = 0$:
\begin{equation}
\tau_s = {h \over a} \left({Y f_\gamma l_\gamma \over 4\pi \Phi}\right)^{1/2}.
\end{equation}
Now the sum $(\lambda_1 + \lambda_2)/a$ can be computed, and solved for
$h/a$:
\begin{equation}
{h(z_s = 0) \over a} = \left({\pi \Phi f_\gamma \over Y l_\gamma}\right)^{1/2}
\left[ {1 - (r_s/a)^2 \over \zeta l_x}\right] \left\{1 - {6 \over \zeta l_x}
\left[ \left({a \over r_s}\right)^2 - 1\right] \left({Y f_\gamma l_\gamma \over
4\pi \Phi}\right)^{1/2}\right\}^{-2}.
\end{equation}
Thus, we can expect the surface to be quite thin near $z_s = 0$ throughout
the interesting regime, {\it i.e.}, when both $l_x$ and $l_\gamma$ are
greater than unity.  However, as the surface accelerates away
from the photon sources, relativistic effects reduce the photon densities
in the moving frame, causing the surface to thicken somewhat.
The condition that the final factor in equation 18
should be near unity is equivalent to $\lambda_2 \ll \lambda_1$.

\section{Summary}

     We have shown that when $L_\gamma/L_x > 0.08$, the pairs that are created
by $\gamma$-ray -- X-ray reactions are squeezed into a screening surface
shaped like an hour-glass at relatively small distances, but which becomes
asymptotically conical at large distances.  As equation 7 demonstrates,
when $l_{\gamma} \gg 1$ and $l_x$ is
large enough to make the surface optically thick to pair production,
the screening surface acquires a significant Compton optical depth at least to
heights $z_s \sim a$.  When that occurs, most
of the incident X-rays are scattered back outside the surface, although
the relativistic bulk motion will beam them within an angle $1/\Gamma$
of the local surface tangent.  As a result, very few X-rays penetrate into
the central region, and the $\gamma$-rays directed within the surface
are free to escape.  This conclusion differs from that of Bednarek (1993),
who found that a compact X-ray ring would lead to $\gamma$-ray absorption,
because he neglected the Compton scattering of X-rays by the pairs.

    On the other hand, most of the energy of the $\gamma$-rays initially
directed into the screening surface is absorbed.  Thus, this mechanism
has the net effect of using a fraction of the $\gamma$-rays (the covering
fraction of the surface around the origin) to preserve the remainder.
When $L_{\gamma}/L_x \sim 1$, the two fractions are comparable.

    Both the escaping $\gamma$-rays and the X-rays are collimated by this
process.  The $\gamma$-rays can only escape in those directions not
covered by the screening surface.  On the other hand, those X-rays initially
directed toward the surface are focussed into directions nearly tangent
to the surface, and also inverse Compton scattered to higher energies.
Therefore, when our line of sight lies nearly parallel to the asymptotic
direction of the screening surface, we should see an X-ray spectrum
which is both quite strong and quite hard.  Thus, objects in which this process
operates can be expected to have strikingly different high energy spectra when
viewed from different directions.

     Because the asymptotic direction of the screening surface changes with
$L_{\gamma}/L_x$, it is possible for fluctuations in $L_\gamma/L_x$ to
strongly modulate the observed X-rays.  When $L_\gamma/L_x$ is small,
our line of sight is likely to lie outside the screening surface, so
we see the unmodified X-ray spectrum; when $L_{\gamma}/L_x$ is large,
the screening surface swings outward, cutting off our view of the X-rays;
for some range of intermediate values, our line of sight will lie close
enough to the direction of the surface that we will see the beamed
and upscattered X-rays.  It is possible that these effects have been observed
in the BL Lac object PKS 2155-304.  Ordinarily its spectral index above a
few keV is in the range 1 -- 2 (Sembay \etal 1993), but in one observation
(Urry \& Mushotzky 1982) the spectral index from 15 keV up to
the sensitivity limit of the instrument, $\simeq 40$ keV, was $\simeq -1.5$! 
Such behavior might be explained if an excursion to especially large
$L_\gamma/L_x$ opened the screening surface wide enough for our line
of sight to be nearly aligned with it.

    In many of the most powerful high energy $\gamma$-ray sources known
there is independent evidence strongly suggesting that much of the
radiation comes from relativistic jets (Hartman \etal 1994; Dermer,
Schlickeiser \& Mastichiadis 1992; Blandford 1993; Sikora, Begelman,
\& Rees 1994). 
Some have argued that the existence of strong high energy $\gamma$-radiation
in these objects is itself evidence for relativistic motion ({\it e.g.}
Zdziarski and Krolik 1993; Dondi \& Ghisellini 1995).  While the mechanism we described here certainly
does not argue against relativistic motion in the source, it does create
a loophole in the arguments for relativistic motion solely on the basis
of high energy compactness.

    Our point of view is that in more realistic pictures of these
sources, {\it both} relativistic motion of the source and optical depth
effects such as the ones we have discussed in this paper may operate.
One might easily imagine, for example, that the toy geometry we have
explored here should, in real sources, be modified to take into account
a relative velocity between the $\gamma$-ray source and the X-ray source
which is very likely relativistic, and may or may not be aligned with
the symmetry axis.  Such relative motion could beam the $\gamma$-rays
away from the X-ray source, so that a smaller fraction of the $\gamma$-rays
are used to produce a shielding wall which is still sufficiently optically
thick to be effective.  This effect could help solve an
otherwise troubling problem in jet models of $\gamma$-ray production in
AGN: that if $\gamma$-rays are produced too close to the center of the
system, pair production on X-rays would lead to a saturated pair cascade
and the production of a much larger X-ray luminosity (through inverse
Compton scattering by the pairs that are produced in the cascade) than
is observed (Ghisellini \& Madau 1995).

   These effects have another significant consequence: there is a collimated
outflow in the screening surface whose luminosity is equal to the absorbed
$\gamma$-ray luminosity, which is likely to be an interesting fraction of
$L_\gamma$.  This outflow, composed of a mixture of electron-positron
pairs and photons of comparable energy, is automatically mildly
relativistic, with a
modest bulk $\Gamma$.  Particularly when $L_\gamma/L_x < 1$ so
that the opening angle is comparatively
small, these outflows have many of the characteristics of the relativistic jets thought to be responsible for
much of the lower-frequency radiation in these objects.

\centerline{Acknowledgments}

   This work was partially supported by NASA Grants NAGW-3129 and NAGW-3156,
Russian Foundation for Fundamental Research Grant 95-02-06063,
and the Space Telescope Science Institute visitors program.  A.F.I. thanks
both the Center for Astrophysical Sciences of Johns Hopkins University and
the Space Telescope Science Institute for their hospitality during his
visit.  We also acknowledge stimulating conversations with Gabriele
Ghisellini and helpful comments from Andrzej Zdziarski.

\clearpage

\centerline{Figure Captions}

\vskip 0.4cm

\noindent Figure 1 \qquad Projection of the screening surface into the $r-z$ plane.  From top to bottom, the curves are for $L_\gamma/L_x = 0.1$, 1.0, 10
and 100.  All assume $\Delta \phi$ is determined by a projection of the
structure into the equatorial plane.  The radius of the X-ray emitting ring
$a$ has been used as the unit of distance.

\noindent Figure 2 \quad $\beta$ as a function of $R = (r^2 + z_s^2)^{1/2}$.
The four curves are for $L_\gamma/L_x = 0.1$, 1, 10, and 100; the
asymptotic value of $\beta$ increases (slowly) with increasing $L_\gamma/L_x$.

\noindent Figure 3 \quad The inertia density $(\mu_e/m_e)\tau_T/l_x$ as a function of
$R = (r^2 + z_s^2)^{1/2}$.  The four curves are for $L_\gamma/L_x =
0.1$, 1, 10, and 100; the
asymptotic value of $\tau_T/l_x $ increases (very slowly) with increasing $L_\gamma/L_x$.
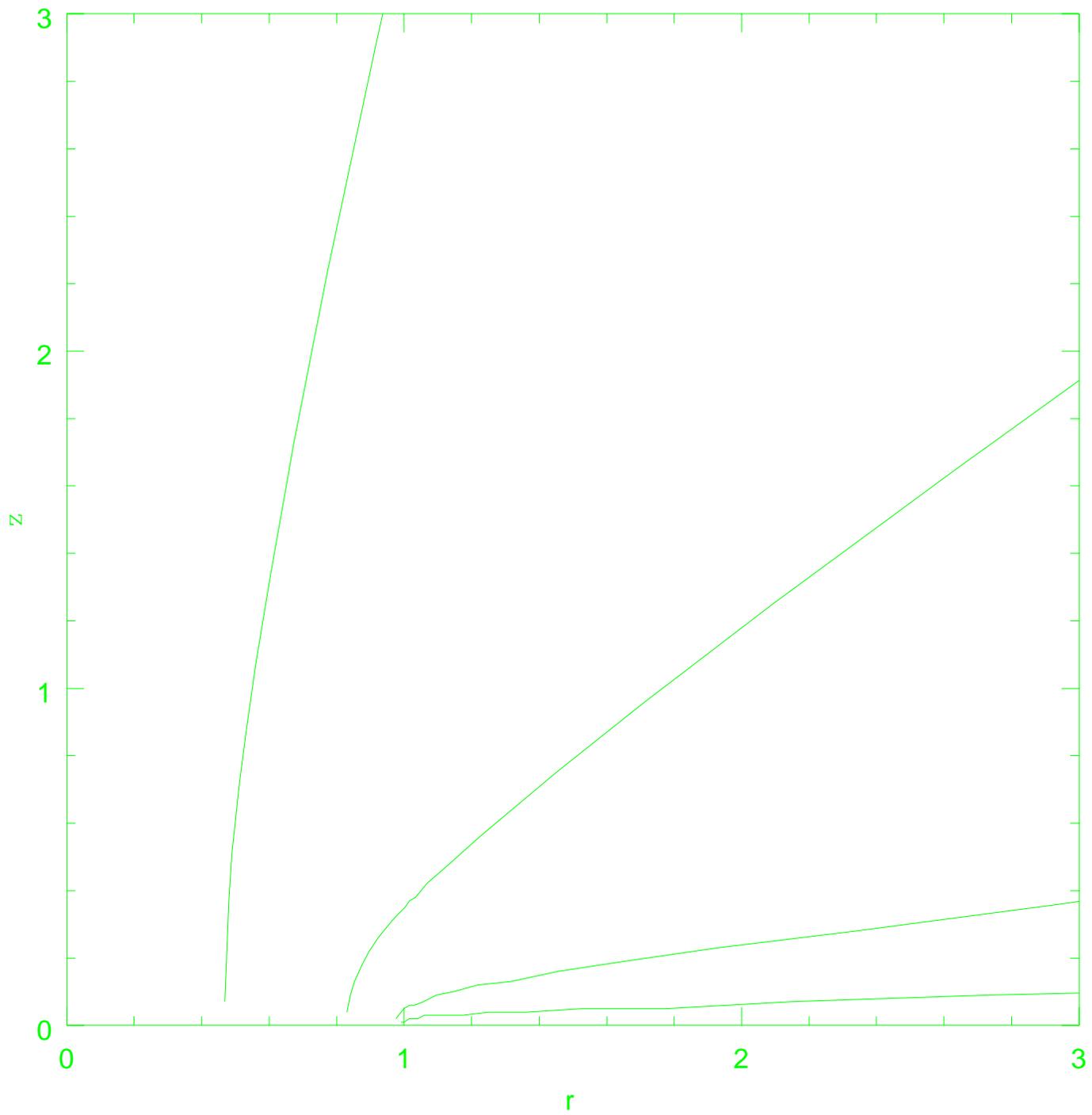
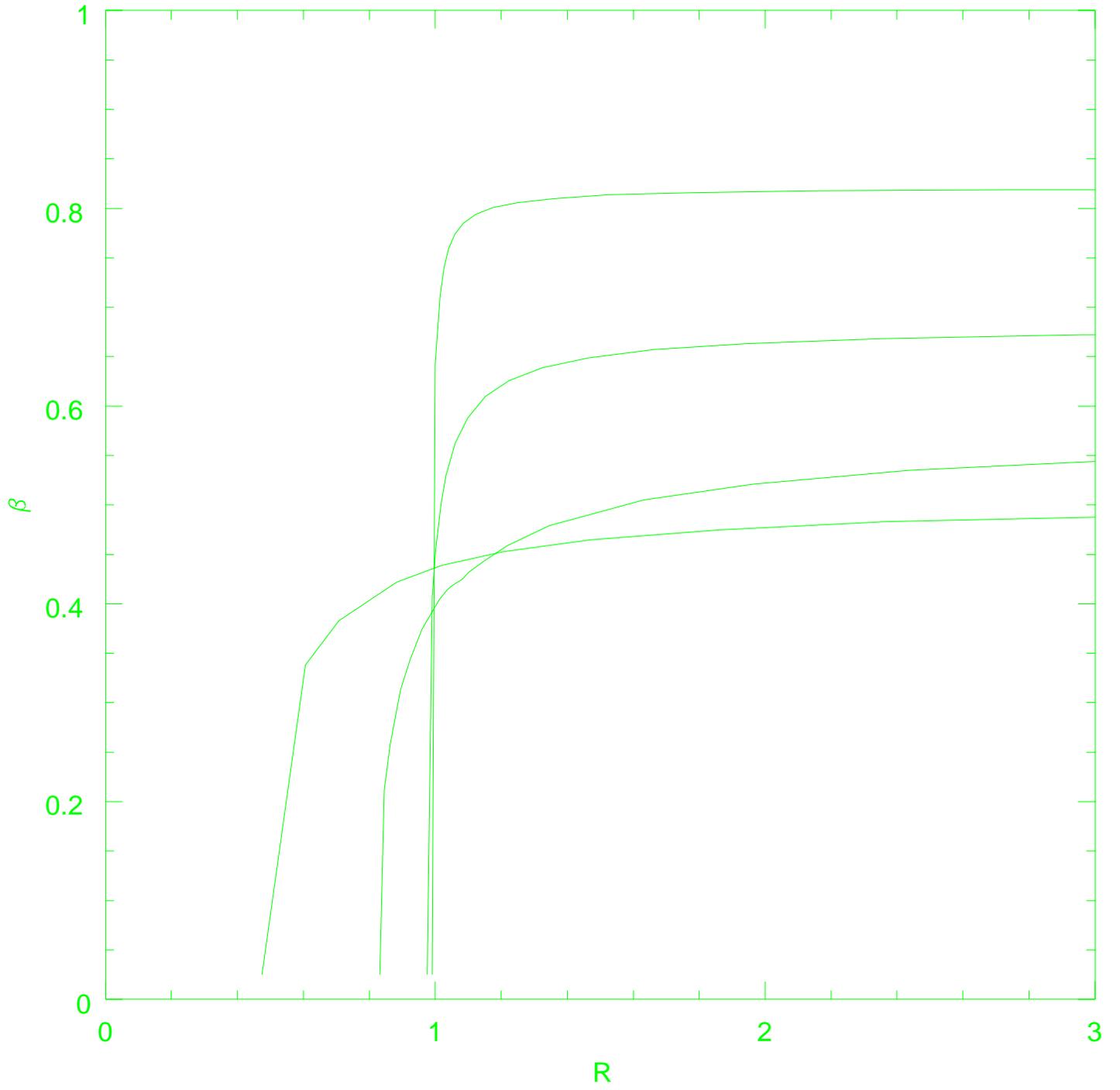
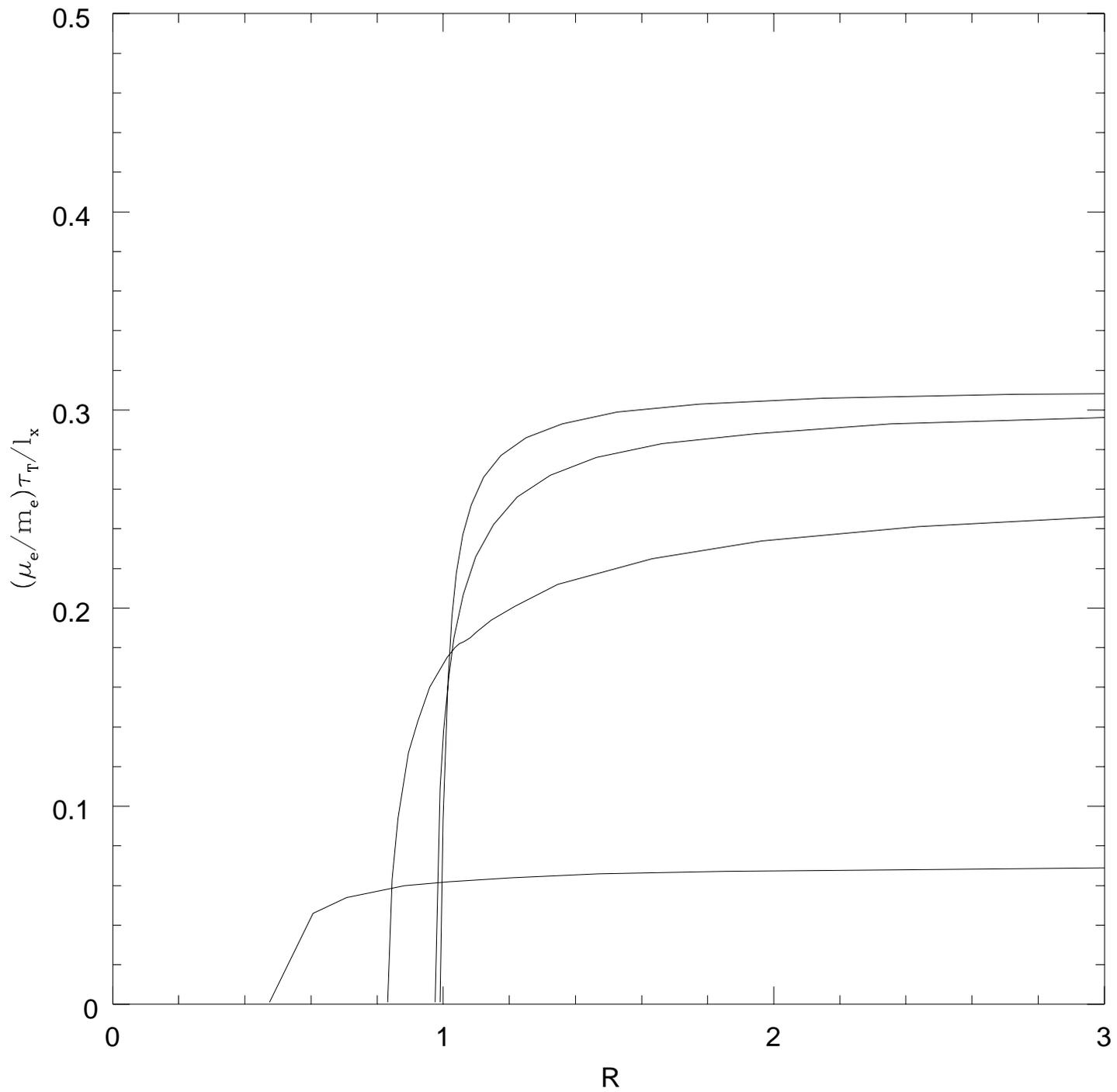
\end{document}